\documentclass[aps,prl, amsmath,amssymb, reprint,]{revtex4-1}
\usepackage{graphicx}
\usepackage{dcolumn}
\usepackage{mathrsfs}
\usepackage{hyperref}
\usepackage{color}
\newcommand{\be}{\begin{equation}} 
\newcommand{\ee}{\end{equation}} 
\newcommand{\bea}{\begin{eqnarray}} 
\newcommand{\eea}{\end{eqnarray}}

\begin{document}
\title{Nonexistence of motility induced phase separation  transition in one dimension}
\author{Indranil Mukherjee, Adarsh Raghu, P.  K. Mohanty}\email{pkmohanty@iiserkol.ac.in}
\affiliation {Department of Physical Sciences, Indian Institute of Science Education and Research Kolkata, Mohanpur, 741246 India.}

\begin{abstract}
We introduce and  study a model of hardcore particles  obeying  run-and-tumble dynamics  on a one-dimensional lattice, where  particles run  in  either +ve or -ve $x$-direction with   an effective speed $v$  and  tumble  (change their direction of  motion)  with  a constant rate $\omega$ when assisted by another particle  from right.   We show that the coarse-grained dynamics of the system  can be mapped  to a beads-in-urn  model called misanthrope process  where  particles  are identified as urns and vacancies as  beads  that hop to a neighbouring urn situated in the direction  opposite to the  current. The hop rate  is  same as  the magnitude of the  particle current; we calculate  it  analytically  for a two-particle system and show  that it  does  not satisfy  the  criteria   required  for a  phase separation  transition.  Nonexistence of phase separation in this model,  where  tumbling dynamics is  rather restricted, necessarily  imply that  motility  induced  phase separation transition  can not occur  in  other models  in  one dimension with unconditional tumbling.  
\end{abstract}\maketitle

An important class of nonequilibrium systems is that of active matter systems (AMS) \cite{ramaswamy_mechanics_2010} where the individual constituents are self-propelled; instances of such systems include bird flocks \cite{ballerini_interaction_2008}, bacterial colonies \cite{hall-stoodley_bacterial_2004}, photophoretic colloidal suspensions \cite{palacci_living_2013} and actin filaments \cite{schaller_polar_2010} etc. They exhibit a number or interesting features like large 
number-fluctuations \cite{ramaswamy_mechanics_2010}, clustering and pattern formation \cite{palacci_living_2013}. A major area of interest in the study of AMS has been the so-called {\it motility-induced phase separation} (MIPS) \cite{cates_motility-induced_2015,tailleur_statistical_2008,thompson_lattice_2011,fily_athermal_2012,redner_structure_2013,bialke_microscopic_2013,levis_clustering_2014,schnitzer_theory_1993} which  refers to spatially  separated  high and low density regimes.  Such aggregation or clustering of particles has been observed  experimentally in many active matter systems \cite{palacci_living_2013}. Relevance of the aggregation  process has also been proposed as a mechanism of  formation   bacterial  biofilms \cite{hall-stoodley_bacterial_2004}, which are sources of infection.

Occurrence of MIPS relies on an argument   that   effective  velocity  of active particles decrease  in crowded or high  density regions  formed  either  by explicit dependence  of local density  or merely by exclusion. Naturally such a slowing down of movement further increases the density of particles and gives rise to a feedback loop allowing the stable high density (liquid-like) regions  to form and coexist  with a low density (gas-like) phase elsewhere.
MIPS has been widely investigated in simulations and  apparent phase separation has been observed. Theoretical investigations of this phenomenon have thus far concentrated on continuum models  \cite{fily_athermal_2012,redner_structure_2013,bialke_microscopic_2013} where  motility parameters, such as particle flux or velocity  are  characterized as functions of the  coarse-grained local density \cite{cates_motility-induced_2015,tailleur_statistical_2008}. Lattice models  of active particles have been  studied in one and two dimensions  numerically  \cite{Golestanian, Whitelam, Soto-2016} with  run and tumble   particles (RTPs). RTPs  move  at a fixed speed along  the direction of their orientation  (a {\it run}) until they {\it tumble} and change their orientation. In one dimension (1D), the two orientations (say, $\pm$)  are usually referred to as   the internal degrees  of the particle ({\it spin}), which flips with a certain rate.  Analytical studies of these lattice  models  are  limited.  Thompson et. al. \cite{thompson_lattice_2011}   have introduced a model of self propelled particles with  RTP dynamics; in 1D.   These  models  exhibit  inhomogeneous  density profiles  when  particle velocities depend on their position. 
Recently Slowman et. al. \cite{slowman_jamming_2016,mallmin_exact_2019} have  obtained  an exact solution for {\it two} RTPs and  found   jamming induced attraction between the particles of the opposite spins, which  indicates  that, for  many particle  systems,
a phase separated state might originate from these  attractive interactions.   Later, Dandekar et. al. \cite{dandekar_hard_2020}  have obtained a mean-field   solution of RTPs  in  1D  which turned out to be  a good approximation  when  tumbling  rate is large. 

An element of surprise in  the formation of a  phase separated state  {\it without} any explicit attractive  interaction  has  generated much  excitement to the study of MIPS and  raised questions about the stability of such states in 1D in absence of  any  explicit  interaction or  spatial  potential. 
Recent works have  added to the doubt by  showing   that  MIPS phase transition in 2D belongs  to  the Ising universality class \cite{MIPS-Ising2D} which  does not have a counterpart in one dimension. In this article we  argue and show explicitly  using   1D lattice models of RTPs that  indeed MIPS transition can not occur in 1D; the inhomogeneous states  observed in numerical simulations and in hydrodynamic models  are only long lived transient states.  

First we  introduce  a  generic model of hardcore RTPs in 1D with a restricted tumbling dynamics and show that its coarse-grained  dynamics can be mapped   to  a beads-in-urn model, namely a misanthrope process \cite{Misanthrope} where  beads hop  to their neighbouring urn, situated in the opposite direction of the particle current, with a rate same  as the  magnitude of current. The  functional form of hop rate is determined from the exact steady state  results of the model with only two RTPs. To determine if  MIPS transition is possible, we use the following criterion.  If a  system of hardcore particles phase separates  as  its density $\rho$ crosses a threshold  $\rho^*$  then  the maximum density at which   it remains homogeneous is $\rho^*.$ Since systems  with homogeneous densities  are  well described   in the grand canonical ensemble (GCE) by  a  unique  chemical potential $\mu$ (or fugacity $z=e^\mu$), we   argue  that  phase separation transition is possible in a system   when  its density  in GCE attains a maximum value $\rho^*={\rm Max}[\rho(z)]$ which is less than unity (the density of a fully occupied lattice). Nonexistence  of MIPS transition in restricted  tumbling  model would imply that MIPS can not occur in any other RTP model  in 1D  where tumbling occurs more frequently.

{\it The restricted tumbling model:} 
We introduce a generic model of RTPs  on an one dimensional periodic lattice with sites  labeled by $i=1,2, \dots L.$  The sites  are either  empty (represented by $\tau_i=0$) or  occupied  by at most one RTP  $\tau_i =\pm$ having orientation (spin) $\pm.$  Particles follow   a run dynamics, 
\begin{eqnarray}
\label{eqn:RTP1}
+0   \mathop{\rightleftharpoons}^{p_+} _{q_+} 0+; ~~
-0   \mathop{\rightleftharpoons}^{p_-}_{q_-} 0-, ~~
\end{eqnarray}
where   RTPs move forward or backward with rates  $p_\pm$  and $q_\pm$  respectively. Along with this, they  can tumble and change their spin
with rate $\omega$ as follows,
\begin{eqnarray}
\label{eqn:RTPw}
+ \pm   \mathop{\to}^{\omega}- \pm ~; ~~~
- \pm   \mathop{\to}^{\omega}+ \pm.
\end{eqnarray}
Tumbling   is  restricted here in  the sense that {\it only}  those  particles  which are  assisted  from right by other particles can tumble their direction.   This  restriction helps us getting  an approximate steady state  of the system without tampering   the main aim: the proposition that a stable  MIPS state  {\it can not}  be sustained  in 1D. Since  frequent tumbling of particles   helps  the system to clear jamming,  a  proof   of nonexistence of  MIPS in  our model necessarily  guarantees  its nonexistence   in any other  model  that has  more liberal tumbling dynamics. Hereafter we refer to  the model following  dynamics  (\ref{eqn:RTP1}) and (\ref{eqn:RTPw}) as  restricted tumbling model (RTM). 

Although   RTM is defined for  generic rates $(p_\pm,q_\pm)$ we 
study the case  $p_\pm= q_\mp$   where the run dynamics  exhibit  
a symmetry  transformation, namely  simultaneous  interchange of parity (left $\rightleftharpoons$ right)   and   spin ($+\rightleftharpoons-$),  that keeps the dynamics invariant. This  symmetry  was present for both run- and tumble-dynamics   in  1D  lattice models studied earlier \cite{slowman_jamming_2016, dandekar_hard_2020}. 
When $p_\pm= q_\mp,$  it is also ensured that in the limit when  lattice spacing  vanishes \cite{Barma},  a single particle dynamics  of   RTM  reduces to   that of a RTP  moving in continuum space  with same  speed  $v=p_-- q_- =  q_+ - p_+$ along  $+$ve and $-$ve  $x$-directions.   Note that, under parity transformation  (left $\rightleftharpoons$ right) 
the tumbling dynamics of our model  is modified as  tumbling now occurs for only those particles which are  assisted  by other particles from left.  But,  for $p_\pm= q_\mp,$  a  left-assisted tumbling dynamics leads  to the same steady state as the right-assisted tumbling. This can be verified  easily   from the exact mapping of  these models to the corresponding beads-in-urn models (see later discussions).

A special case of RTM with $p_+=\alpha= q_-,p_-=0= q_+$  and unrestricted tumbling  dynamics  $+\mathop{\rightleftharpoons}\limits^{\omega} _{\omega}-$ was  studied  earlier by Slowman et. al. \cite{slowman_jamming_2016} 
and an exact  steady state solution was obtained for a  system of two RTPs. It turned out that  these two particles   experience an effective attractive interaction in the steady state when their spins are opposite; it is envisaged  that this attraction   might be the source of MIPS states observed in corresponding hydrodynamic models. In comparison,  in Eq. (\ref{eqn:RTPw})  we have  dropped  one  of  the  transition  $+ 0   \mathop{\rightleftharpoons}\limits^{\omega} _{\omega}- 0;$ as a consequence, particles do not  tumble if  they are not assisted by a right neighbour.

\begin{figure}
\includegraphics[width= 3.5 in]{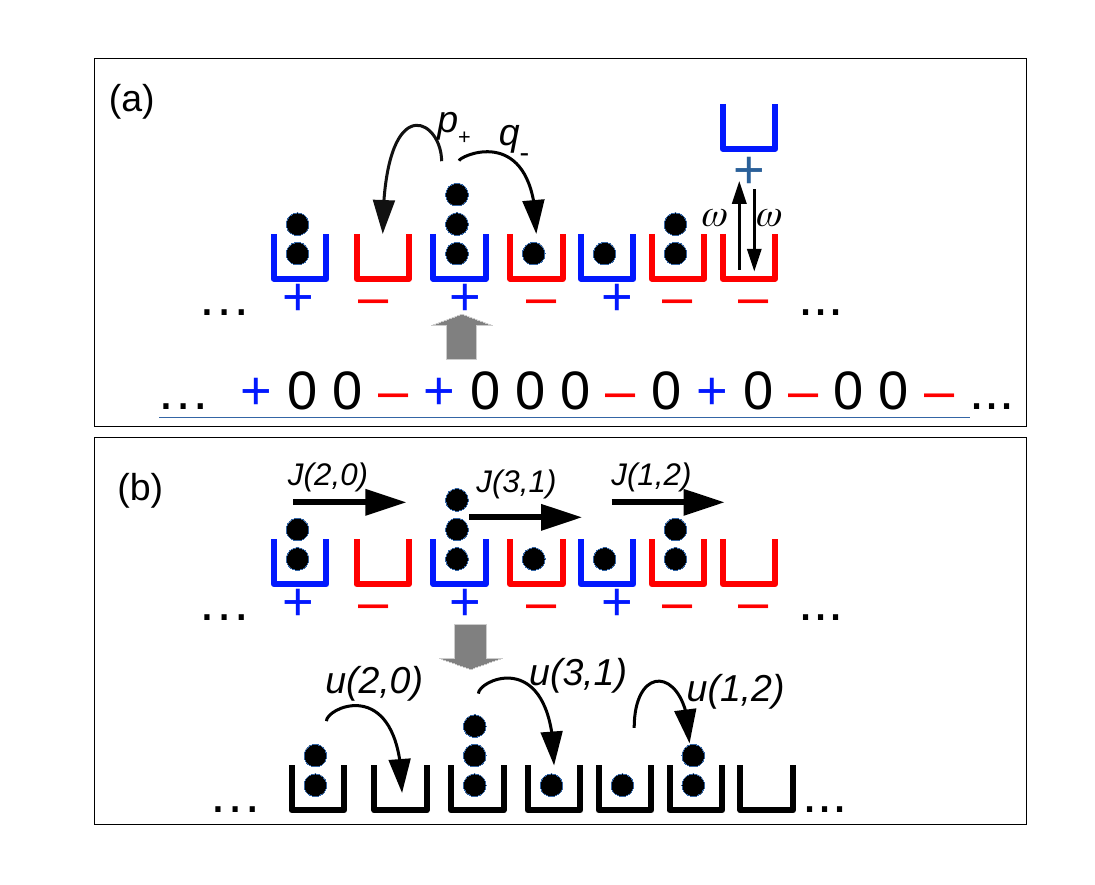}\vspace*{-.8 cm}
\caption{(a) Mapping lattice model of   RTPs to an urn model. (b) Effective coarse-grained dynamics:  hop rate of a bead $u(m_k,m_{k+1})$  from  urn $k$ (with $m_k$ particles) to $k+1$  (with $m_{k+1}$ particles)  is assumed to be same as the local bead current  $J(m_k,m_{k+1})$ averaged over   internal degrees $\sigma_k,\sigma_{k+1}.$ \label{Fig1}}
\end{figure}

{\it Mapping  to beads-in-urn model:} Any microscopic configurations $\{ \tau_i\}$ of RTM  can be  viewed as urns containing beads --each particle is an urn that contains 
beads   which are   uninterrupted  sequence of 0s (vacancies) 
to the right of the  particle (as described in Fig. \ref{Fig1}(a)).  The spin $\pm$ of the particle is termed as the internal degree  of the  urn. Thus   we have a beads-in-urn model 
of $N$ urns indexed by $k=1,2,\dots N,$ each carrying an internal degree  $\sigma_k=\pm$ and  $m_k= 0,1,2\dots$ beads. 
The dynamics (\ref{eqn:RTP1}) and  (\ref{eqn:RTPw}) now  translate to hopping of a bead from  urn $k$  to $k+1$ ($k-1$)  with rate $q_{\sigma_{k+1}}$  ($p_{\sigma_k}$),  and flipping  of internal degrees $\sigma_k \to -\sigma_k$  with rate $\omega  \delta_{m_k,0}.$ The total number of beads $\sum_{k=1}^N m_k = L-N\equiv M$ is conserved  by the dynamics. 
Like particle density $\rho= \frac{N}{L},$ the  bead density
$\eta = \frac{M}{N} = \frac{1-\rho}{\rho}$ is also conserved. 

Note that in this  beads-in-urn model  the  internal degrees of the urns  can flip  only when they are empty; this restriction forces  $k$-th  urn either to  transfer  a bead (when $m_k>0$){\it or}  to change the internal degrees (when  $m_k=0$) and help us  getting an exact steady state. It is easy to see that  a left-assisted tumbling dynamics with  same rate $\omega$  will also map  to  the {\it same} beads-in-urn dynamics when  particles  are identified as urns   containing  number of beads  same as  the  
consecutive vacancies  to their left  and   the hope rates are 
 $p_\pm= q_\mp.$
 
The mapping  of   RTM   to beads-in-urn model  is  exact but its  steady-state could not be obtained analytically. We proceed  to develop a coarse-grained picture. In the steady state of the urn model,  the local   bead current $J$
(summed  over  $\pm$ degrees) effectively  transports the beads from one urn to  its  neighbour situated  along the direction of total current. Since hop-rates ($q_{\sigma_{k+1}},p_{\sigma_k}$) in the original beads-in-urn model were dependent on  spins of neighbouring  urns it is  expected that the  local bead current must depend on the number of beads present in neighbouring urns, i.e.  $J \equiv J(m_k,m_{k+1}).$  This current can be set  as the effective  hop-rate   of  a coarse-grained  model  where urns  lose their  internal degrees and a single bead hops from  urn $k$  to $(k+1)$  with rate  $u(m_k,m_{k+1})=J(m_k,m_{k+1});$   rightward hopping ($k$  to $(k+1)$) is considered assuming that the current  is flowing in $+$ve $x$-direction. Thus,
in this coarse-grained picture (see Fig \ref{Fig1}(b)),  all urns are equivalent (as  they lose their  internal degrees) and the hop-rate  depends on the number  of beads present in  the departure and the arrival urn; such a process is called a misanthrope process (MAP)\cite{Misanthrope}.

In fact, mapping of hardcore particle systems to  urn model  with an exact or effective coarse-grained dynamics, similar to the dynamics of a {\em zero range process} (ZRP) \cite{ZRP-Evans} are  quite reliable   and have  helped researchers \cite{Kafri}  earlier to establish  non-existence  of  phase separation   transition  in  certain  lattice models \cite{AHR} where  rigorous  numerical  simulations  have exhibited apparent phase separated states. It also helped  in predicting  {\em true} phase separation  transition in  many other models \cite{Kafri, PK-David, PK-Kundu}.  In contrast, mapping  to  that of  misanthrope process,  that we introduce  here, provides a better coarse-grained picture as steady-state correlation  between neighbouring urns are  retained here.  
  
The bead-current $J(m_k,m_{k+1})$ flowing across  the urns can be computed  from numerical simulations (will be discussed later), but that does not  help us  to compute  $\rho(z)$  in  grand canonical ensemble. To calculate $\rho(z)$ we  need functional form of  $J(m_1,m_2)$ which can be calculated exactly   using  matrix product  ansatz (MPA)\cite{MPA}  for a system of  two urns  containing $M$ number of beads (i.e.,  $L=M+2$), each one  following   the dynamics described in Fig. \ref{Fig1}(a).

For urn models, a  matrix product steady state (MPSS)  can be obtained   following Ref. \cite{Amit}. 
We now  consider RTM model,  which is mapped exactly to the urn model described in Fig. \ref{Fig1}(a). The steady state probability of  a generic configuration $\{\sigma_km_k\},$  where $k^{th}$ urn (spin $\sigma_k$)   has $m_k$ beads,  is  given by  a matrix product ansatz, 
\bea \label{ansatz}
P(\{\sigma_km_k\}) \sim  Tr[ \prod_{k=1}^N X_{\sigma_k}(m_k)]\delta \left(\sum_{k=1}^N m_k -M\right)
\eea
where  matrix $X_{\sigma_k}(m_k)$ represents the $k^{th}$ urn having  internal degree $\sigma_k$  and $m_k$ beads.  The $\delta$-function  here ensures   that the total  number of  beads $M$ are conserved. 
These matrices are constrained to follow  a matrix algebra  so that   $P(\{\sigma_km_k\})$ defined above must satisfy the steady state condition $\frac{dP}{dt} =0$ for the dynamics  in Fig \ref{Fig1}(a). We find  (see Appendix) that for $N=2,$  matrices $X_{\sigma}(m)$  have a  $2\times 2$ representation (for any  $\omega>0$), 
\be\label{eq:matrix}
\hspace*{-.3 cm} X_+(m)=  \begin{bmatrix} 1 & 0\\1 & 0 \end{bmatrix}, 
X_-(m)= \gamma^m \begin{bmatrix} 0 & 1\\0 & 1 \end{bmatrix};~ \gamma = \frac{p_+ + q_-}{p_-+q_+}.
\ee
The steady state probabilities of two urns containing $m_1,$ $m_2$ beads are then, $P_{\sigma_1 \sigma_2}(m_1,m_2)=\frac{1}{Q_M}  Tr[ X_{\sigma_1}(m_1) X_{\sigma_2} (m_2)] \delta(m_1+m_2-M)$ where 
 $Q_M={\sum _{\sigma_1, \sigma_2} \sum_{m_1=0}^{M} Tr[ X_{\sigma_1}(m_1) X_{\sigma_2} (M-m_1)] }.$ 
Explicitly,
\be \label{eq:PN=2}
P_{\sigma_1 \sigma_2}(m_1,m_2)=\frac{1}{Q_M} \gamma^{ \frac12(1-\sigma_1) m_1 + \frac12 (1-\sigma_2) m_2},
\ee
with $m_2= M-m_1.$
Thus, the average  local current carried by the beads when the two  urns have $(m_1, m_2)$ particles is 
\bea 
J(m_1,m_2)&=&\sum_{\sigma_1,\sigma_2}P_{\sigma_1\sigma_2}(m_1,m_2)(q_{\sigma_1} - p_{\sigma_2})\cr
 &=& \frac{1}{Q_{m_1+m_2}} [   (q_+-p_+) + (q_+-p_-) \gamma^{m_1} \cr
&+& (q_--p_+) \gamma^{m_2} + (q_--p_-) \gamma^{m_1+m_2}] \nonumber
\eea 
For RTPs, which need   to satisfy the condition $p_\pm= q_\mp,$ 
\be \label{eq:JM}
 J(m_1,m_2)  =v\frac{1 - \gamma^{m_1+m_2}}{Q_{m_1+m_2}},
\ee 
where  $v=p_--q_-= q_+-p_+$ and $\gamma = \frac{ p_+}{p_-}$ (as in Eq. (\ref{eq:matrix})). Note that  $J(m_1,m_2)$ depends  only on the sum  of its arguments, i.e., $J(m_1,m_2) \equiv J(m_1+m_2).$   We will now
set  $J(m_1+m_2)$ as   the hop-rate of beads  in the coarse-grained model, i. e., $u(m_k, m_{k+1})=  J(m_k+ m_{k+1})$.
This urn model is a  misanthrope process where hop-rate is a function of  total number of beads present in   the  departure and the  arrival site.   It turns out that  the steady state  of this specific 
misanthrope process has a factorized form, 
\be 
P(\{m_k\}) \sim  \prod_{k=1}^N f(m_k) ~{\rm  with}~  f(m) =  \prod_{n=1}^m \frac{u(1, n-1)}{u(n,0)}=1.
\nonumber
\ee
The   grand partition function  with a fugacity $y$ that controls the total number of beads $M\sum_{k=1}^N m_k$ is 
\bea
{\cal Q}_N(y) &=& \sum_{\{m_k\}}  P(\{m_k\}) y^{m_k} = F(y)^N;\cr
F(y) &=& \sum_{m} f(m) y^{m}=\frac{1}{1-y}.
\eea
In RTM, both $N,M= \sum_{k=1}^N m_k$ vary  keeping the system size  $L$ fixed. To account for that  
we  introduce  another  fugacity $z,$  so that  the new partition function  is, 
\be
Z(z,y) =  \sum_{N=0}^\infty {\cal Q}_N(y) z^N 
= \frac{1}{1-z F(y)} \label{eq:Z} 
\ee
which gives rise to $\langle N\rangle  =z\frac{\partial}{\partial z} \ln Z(z,y)$ and $\langle M\rangle=y\frac{\partial}{\partial y} \ln Z(z,y).$ We now set  $\langle N\rangle + \langle M\rangle \equiv L$  to  obtain $z$ in terms of $y,$ $z= \frac{L}{(1+L) F(y) +y F'(y)}.$ Then,    
\be \label{eq:rho}
\rho(y)\equiv \frac{ \langle N\rangle}{L}= \frac{F(y)}{F(y)+y F'(y)} =1-y.
\ee
The maximum value of the  RTP density, obtained  when $y\to 0,$
$\rho^*=1$ (fully occupied lattice). Thus  
the system remains  homogeneous  for any density $0<\rho<1$   and it {\it can not} phase separate 
(following the criterion we discussed).
\begin{figure}
\includegraphics[width= 4.22cm]{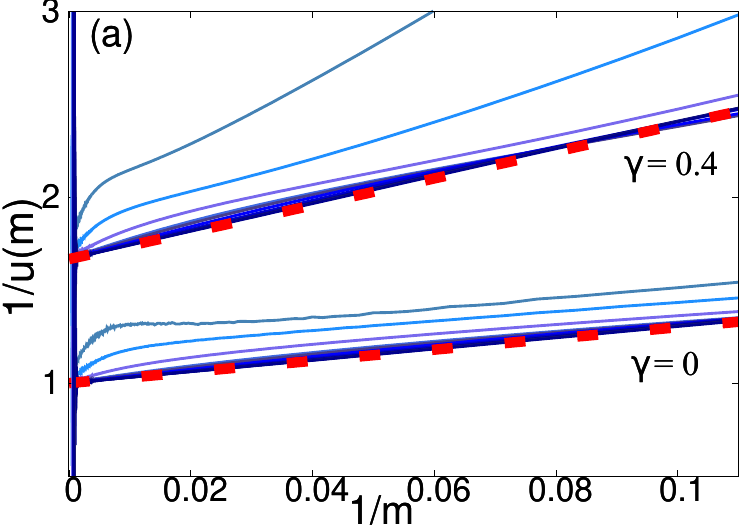}\includegraphics[width= 4.24cm]{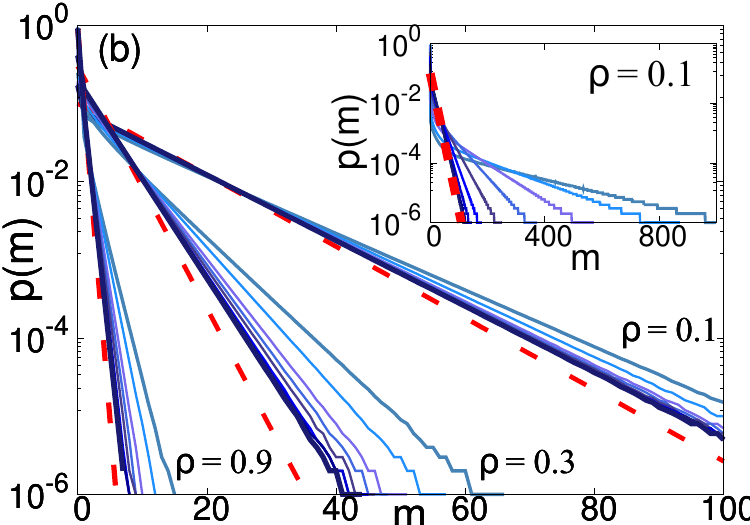}\caption{ \label{u_m} Simulation of RTM model   with dynamics (\ref{eqn:RTP1}) and (\ref{eqn:RTPw})  (equivalently an urn model described in Fig. \ref{Fig1}(a)).
(a) Hop rate $u(m)^{-1}$ obtained from numerical simulations (solid line)   for   $\gamma=0, 0.4$  and   $\omega=0.005$ to $1$ (top to bottom)  are compared  with  Eq. (\ref{eq:J_largeM}) (dashed line)  when $\rho=0.02.$ All the curves approach linearly to  the asymptotic value $(1-\gamma)^{-1},$   as predicted. (b) Marginal distribution $p(m)$ of the  separation $m$  are compared for $\gamma=0$ and  $\rho=0.1,0.3, 0.9$ in semi-log scale.  Solid lines  (results from simulations for $\omega= 0.2$ to $10$ (right to left) are shown  along with   dashed lines, $\rho(y)^m$ with $y=1-\rho$ obtained from  coarse-grained description of the model.  The inset shows the same for $\rho=0.1$  but smaller  $\omega= 0.005$ to $1$ (right to left). In all cases  $p(m)$ shows exponential behaviour; but for small $\omega,$ $y$ differs  substantially from the predicted value $(1-\rho)$.  Here, $p_+ =\gamma= q_-,$  $p_-=1=q_+,$ $L=10^4.$  In each case, statistical averaging is done for more than  $10^7$ samples.}
\end{figure}

The above argument is based on a coarse-grained picture where the hop rate $u(m,n)\equiv u(m+n)$ is  taken  same as the average   local current of beads. 
In the following  we employ a method to calculate  $J(.)$ numerically  from Monte Carlo  simulations of the model and 
compare it with Eq. (\ref{eq:JM}).  

To  simulate  the dynamics  we must set   $p_\pm= q_\mp$ required   for  the system to have a valid RTP dynamics, which gives  $\gamma= \frac{p_+}{p_-}$ in Eq. \ref{eq:matrix}. Without  loss of generality  we can set   $p_-=1=q_+,$ by choosing a suitable time unit; then,  $p_+ = q_-= \gamma$ and  the  speed  of RTPs $v= q_+-p_+ =1-\gamma.$  We also consider  $\gamma\le1$   ($\gamma>1$ case   can be explored directly by  using left/right and $+/-$ symmetry). From  Eq. (\ref{eq:JM}), $J(m) = \frac{v}{Q_m} (1- \gamma^m),$  which has an asymptotic form  (for large $m),$
\be\label{eq:J_largeM}
J(m)\equiv u(m) \simeq m\frac{1-\gamma}{m +c}; c=\frac{3-\gamma}{1-\gamma}.
\ee
This implies  that   $u(m)^{-1}$ is a linear function of $m^{-1}$ with slope $c(1-\gamma)^{-1}$ and $y$-intercept $(1-\gamma)^{-1},$  which we  verify 
from the  Monte Carlo simulations of  the  urn model (Fig \ref{Fig1}(a)).   For a given value of $\gamma, \rho, \omega$ 
first we  allow the system to relax   for a long time starting from a  random initial configuration. The system may take a very long time to reach a true phase separated state  when it exists, but the  hoping dynamics in the coarsening regime given by $u(m_1,m_2)=J(m_1,m_2)$ can  predict, well in advance, if the system is  approaching  towards  a inhomogeneous (MIPS)  or a homogeneous state.

In the coarsening regime we consider a large  time interval  and  calculate  $(F_r(m_1+m_2),  F_l(m_1+m_2)),$  the number  of times beads  move  to  (right, left) when the  departure and arrival urns have exactly $m_1$ and $m_2$ beads respectively (internal degree of the urns are ignored). Also,  we keep track of $F(m_1+m_2),$  the   number  of jump-events attempted during that interval. Clearly,  $u(m)=  (F_r(m) -F_l(m))/F(m).$ In Fig. \ref{u_m}(a) we  plot $u(m)^{-1}$  versus $m^{-1}$   for $\gamma=0,0.4, \rho=0.02$ and  $\omega =0.005$  to $1;$ in all cases, $u(m)^{-1}$ is  found  to be linear  for large $m$ 
as expected  from Eq. (\ref{eq:J_largeM}).  The  $y$- intercepts also  approach   to  the known value $(1-\gamma)^{-1}$  but the slopes differ a bit.  Further, in  Fig. \ref{u_m}(b) we  plot the marginal  distribution  $p(m)$ of number beads $m$ for $\gamma=0, \rho=0.1, 0.3, 0.9,$ $\omega=0.2$ to $10$.  The dashed line   corresponds  to  the  theoretical curve obtained from the coarse-grained picture: $p(m) = y^m f(m)/F(y) = \rho y^m$ where $y= 1-\rho.$  In all cases, as shown  Fig. \ref{u_m}(b), $p(m)$ exhibits exponential  distributions that match very well with the prediction  when  $\omega$ is large.  As $\omega \to 0$ the  exponential feature remains  persistent but the value of $y$ differs   
substantially from the theoretical value $1-\rho.$ This is   because ergodicity is broken at $\omega=0;$  the system there falls into  one of the fully jammed  (or absorbing) configuration and remains there.

Essentially, the  coarse-grained  picture  turns out to be a  good description  of the RTP model as  $p(m)$ decays  exponentially  for large $m$ as predicted - rest of the details
are  less relevant  because an  exponential form of $p(m)$  is enough  to  assure that the fugacity  in GCE can {\it always} be tuned to secure  any desired  particle density $0<\rho<1.$ Such a system can not support any  stable MIPS phase  and settles  to form a homogeneous density  profile  for   all $\omega>0,\gamma\ge0.$  

The above conclusion can also be obtained from using an approximate matrix product steady  state (MPSS).  Matrix representations   (\ref{eq:matrix}),  that provides exact MPSS  exclusively for $N=2,$ are also  excellent  approximations for   larger $N$ (justified in  the Appendix). 
With  these matrices, for $N>2,$ the grand  partition function 
$Z(z,y)$ and density $\rho(y)$  are given by Eqs. (\ref{eq:Zzy_app}) and (\ref{eq:rhoy_app}) respectively, 
\bea
Z(z,y)
&=& \frac{1}{1- z F(y)};~ F(y)=\frac{1}{1-y} +  \frac{1}{1-\gamma y}\cr
{\rm and}~\rho(y)&=&\frac{(1-y)(1- \gamma y) (2-y- \gamma y)}{ (1- \gamma y)^2 + (1-y)^2}.
\eea
Clearly,  the maximum  density that  can be achieved  in GCE  by tuning  $y$ is $\rho^*=1$ (when $y=0$)  and thus,  this RTP model {\it can not} undergo  a phase separation transition at any $\rho<1.$   One can   safely extend  these results  for restricted tumbling dynamics to  other RTP models where  tumbling occurs  more frequently; this is  because   tumbling  is  generally  detrimental to the stability of MIPS. Our conclusions are consistent with  the recent results \cite{MIPS-Ising2D}  that  MIPS transition in 2D   belongs to  the  Ising universality class that  does not have an one dimensional  analogue.

In summary, we show that  phase separation of {\it free}  hardcore-RTPs
with constant  run and tumble rates is not possible in 1D. One may however add some crucial features which  are known to enhance or freshly produce phase separated states of passive particles, like invoking explicit attractive interaction \cite{Kafri} {\it or}  making  tumbling  rates  to decrease with  $L$ (so that it vanishes in the thermodynamic limit) \cite{Urna} {\it or} explicitly  forcing  the run dynamics  to depend on  (and reduce substantially  with increase of) local particle density \cite{ZRP-Evans} {\it or}  adding   impurities \cite{ZRPdefect}. Then a phase separation transition  may occur,   but will it   keep its charm  and glory to be identified as the {\it motility induced} phase separation, particulary    when   the transition is anyway expected for similar system of passive particles (without   motility)? 
Recently  Kourbane-Houssene et. al. \cite{Kourbane-Houssene} have  introduced a RTP model where the    difference of  run-rates   (or effective velocity) are taken  proportional to $\frac1L$ and the tumbling rate is proportional to $\frac1{L^2}$ (downplayed by a factor $1/L$ compared to the run rates);     using an  exact  coarse-grained  hydrodynamic  description  they show that a homogeneous phase in 1D   loses its stability in  certain parameter regimes. Another way  might be  to use  strongly biased   tumbling  rates where, say,   $+\to -$  occurs much more frequently than $-\to +.$  In this case a phase separation transition  occurs  \cite{Urna}  when  $q_\pm=0,$ 
where  the dynamics of  RTM reduces to  that of a  two species exclusion  process \cite{Urna-PK}. Its extension to small $q_\pm\simeq0,$  is a RTP model  (having a good continuum limit) and it is reasonable to assume that the phase separation  features   may also  survive  there. Yet another possibility is to introduce defects. Recent  studies \cite{Amit-2}  have shown  that  a jammed phase does exist in  RTM  like models  with defects. More investigations are  required  in all these directions  to confirm if RTP models in 1D can  phase separate. 
\vspace*{.5 cm}

\centerline {\bf APPENDIX} \vspace*{.15 cm}
The  dynamics  (\ref{eqn:RTP1}) and   (\ref{eqn:RTPw})  of RTM  can  be mapped  exactly  to an  urn model  described in Fig. \ref{Fig1}(a)  where beads  hop  from site $k$ to site $k+1$ (or site $k-1$) with rates $q_{\sigma_{k+1}}$  (or $p_{\sigma_{k}}$) respectively. 
The  probability density  of a  generic configuration $ \{ \sigma_km_k\}$ evolves  following the  Master equation, 
\bea
&&\frac{d}{dt} P( \dots,\sigma_{k-1}m_{k-1}, \sigma_km_k,\sigma_{k+1}m_{k+1},\dots) \cr
&&~~ = - (p_{\sigma_k} + q_{\sigma_{k+1}})  P( \dots,\sigma_{k-1}m_{k-1}, \sigma_km_k,\sigma_{k+1}m_{k+1},\dots)\cr
&&~~~~~  + q_{\sigma_k} P( \dots, \sigma_{k-1}m_{k-1}+1, \sigma_k m_k-1,\sigma_{k+1}m_{k+1},\dots)\cr
&&~~~~~ + p_{\sigma_{k+1}} P( \dots,\sigma_{k-1}m_{k-1}, \sigma_k {m_k-1},\sigma_{k+1}m_{k+1}+1,\dots)\cr
&&~~~~~ -\omega\delta_{m_k,0} P( \dots,\sigma_{k-1}m_{k-1}, \sigma_km_k,\sigma_{k+1}m_{k+1},\dots)\cr
&&~~~~~ +\omega \delta_{m_k,0} P( \dots,\sigma_{k-1}m_{k-1}, -\sigma_km_k,\sigma_{k+1}m_{k+1},\dots) 
\eea
where first  three terms  in the right hand side corresponds to the run dynamics   and the rest  describes tumbling   at a generic site $k.$ In the steady state $\frac{d}{dt}P(\{\sigma_km_k\})$ must vanish; this, along with  the matrix product  ansatz (\ref{ansatz}) leads  to 
$\sum_{k=1}^{N} {\rm Tr }[ H^R_{k}  +H^{T}_{k}] =0,$ where $H^R_{k}$ and $H^{T}_{k}$ correspond to the run  and the tumble  dynamics  respectively,   
\bea \label{eq:cancel}
&&H^R_{k}=- (p_{\sigma_k} + q_{\sigma_{k+1}})  X_{\sigma_{k-1}} (m_{k-1}) X_{\sigma_{k}} (m_{k})X_{\sigma_{k+1}} (m_{k+1})\cr && ~~~~
+q_{\sigma_{k}}  X_{\sigma_{k-1}} (m_{k-1}+1) X_{\sigma_{k}} (m_{k}-1)X_{\sigma_{k+1}} (m_{k+1})
\cr && ~~~~ + p_{\sigma_{k+1}} X_{\sigma_{k-1}} (m_{k-1}) X_{\sigma_{k}} (m_{k}-1)X_{\sigma_{k+1}} (m_{k+1}+1),\cr\cr
&&{\rm and}~H^{T}_{k}=\omega [ X_{-\sigma_{k}} (0)-  X_{\sigma_{k}} (0)]X_{\sigma_{k+1}} (m_{k+1}).
\eea
We now introduce some suitable choice of   auxiliary matrices $\tilde X_{\sigma_k, \sigma_{k+1}} (m_{k}, m_{k+1}),$  yet to be determined  along with $X_{\sigma_k}(m_k),$   so that both  $\sum_k H^R_{k}$ and  $\sum_k  H^{T}_{k}$ vanish separately;  one such  cancellation scheme  for $H^R_{k}$ is,
\bea
H^{R}_{k} &=&\tilde X_{\sigma_{k-1},\sigma_{k} } (m_{k-1}, m_{k}) X_{\sigma_{k+1}} (m_{k+1})\cr
&-& X_{\sigma_{k-1}} (m_{k-1}) \tilde X_{\sigma_{k},\sigma_{k+1} } (m_{k},m_{k+1} ).
\eea
We find that a choice $\tilde X_{\sigma, \sigma'} (m,n) = h_{\sigma \sigma'} X_{\sigma} (m) X_{\sigma'} (n)$   with some  scalar parameter $h_{\sigma\sigma'}$ does satisfy the steady state condition with $2\times 2$ matrices
\be\label{eq:matrix-A}
X_+(m)=  \begin{bmatrix} 1 & 0\\1 & 0 \end{bmatrix}, 
X_-(m)= \gamma^m \begin{bmatrix} 0 & 1\\0 & 1 \end{bmatrix},
\ee
when $\gamma = \frac{p_+ + q_-}{p_-+q_+},$ $h_{+-}= 0=h_{-+}$ and 
\be  \label{eq:h}
 h_{++}=  h_{--}=\left\{  \begin{array}{ll}
                               q_\sigma(1- \gamma^\sigma) &  m>0,n>0\\ 0  & {\rm else}
                             \end{array}
\right..
\ee
These matrices  also satisfy the condition $\sum_k  {\rm Tr}[H^{T}_{k}]=0$ set by the tumbling dynamics because $X_\sigma(0) X_{\sigma'}(m)=   X_{\sigma'}(m)$ for all $\sigma,\sigma', m.$ 
The only troubling part  is that $h_{\sigma\sigma}$s   depend implicitly  on  $m,n$  violating the assumption that they are constants. This implicit dependence of $h_{++}$  and $h_{--}$  on $m,n$  drops out  when 
(i) $q_\pm =0$   (all particles move in the  same direction), 
(ii)$\gamma =1$ (which sets  the speed of RTPs $v= 1-\gamma=0$  when $p_\pm=q_\mp$).
In both cases we have an exact MPSS, but neither of these cases  constitutes the scenario of MIPS.  Yet another case is $N=2$ where matrices   given by Eq. (\ref {eq:matrix-A}) leads to an exact MPSS. This is because the cancellation scheme in Eq.  (\ref{eq:cancel}) acts on  product  of {\it three}  consecutive  matrices which are   not present when $N=2;$ thus, one can 
make $h_{\sigma\sigma'}$ independent of $m,n$  by setting 
safely $h_{\sigma\sigma'} =0$ for all $\sigma, \sigma'.$ 
Steady state  probabilities  for $N=2$ is given by Eq. (\ref{eq:PN=2}). 
\begin{figure}
 \includegraphics[width=1.6in]{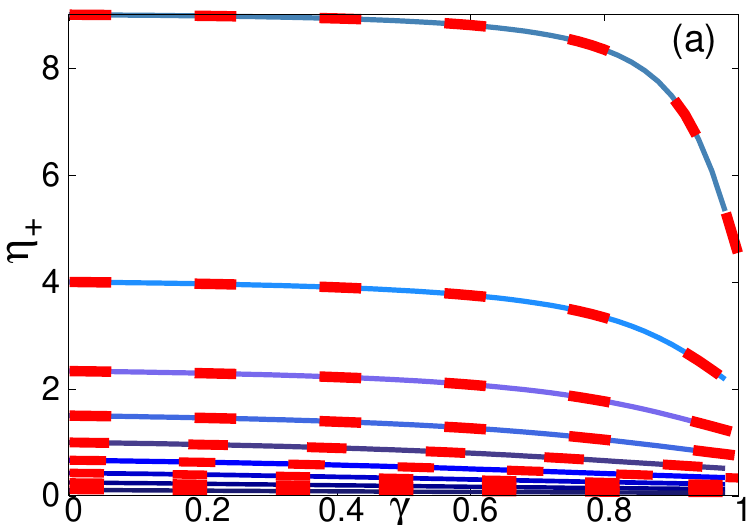}\includegraphics[width=1.6in]{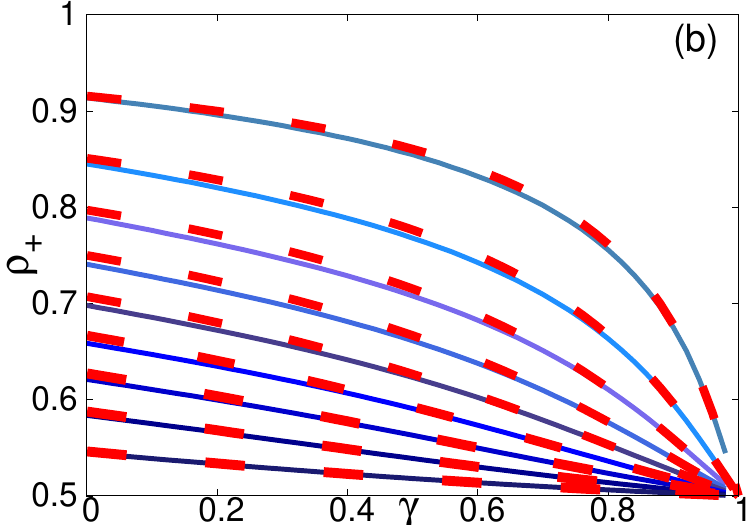}
 \caption{ (a) $\eta_+$,  the  density of beads  in $+$ urn and (b) $\rho_+$,  the  fraction of  $+$ urns   are shown as a function of  $\gamma$   for different $\rho= 0.1$ to $0.9$ (top to bottom). Data  from Monte Carlo simulations (solid lines) of RTM  model described in Fig. \ref{Fig1}(a), averaged over  $10^7$ samples are compared with Eqs. (\ref{eq:etapm}) (dashed line). Other parameters  are $L=10^3$, $p_+=\gamma= q_-,$ $p_- =1=q_+$ and $\omega=1.$ 
 \label{Fig:approx}}
\end{figure}

Now we proceed  for larger $N$ and get an approximate  MPSS
while  dependence of $h_{\sigma\sigma'}$    
on $m,n$   are ignored  and  both $h_{++}$ and $h_{--}$ are  taken as  $q_\sigma(1- \gamma^\sigma)~~\forall m,n\ge0.$  
We will see  that  the  matrices  (\ref{eq:matrix-A})  provide a MPSS which are  an excellent approximation to the  exact ones.  
The canonical partition function  of the system is  
\bea
Q_{M,N}=  \sum_{\{\sigma_k m_k\}} Tr\left[   \prod_{k=1}^N  X_{\sigma_k}(m_k)\right] \delta \left(  \sum_{k=1}^N m_k -M\right)\nonumber
\eea
and the  grand partition function, with fugacities $z,y$ associated with $N,M,$ is  
\bea \label{eq:Zzy_app}
Z(z,y) = \sum_{M=0, N=0}^\infty z^N y^M  Q_{M,N} = \sum_{N=0}^\infty z^N F(y)^N\cr
F(y)= \sum_{\sigma=\pm} \sum_{m=0}^\infty y^m Tr[ X_\sigma(m)] = \frac{1}{1-y} +  \frac{1}{1-\gamma y}.
\eea
Note that $F(y)^N$ acts  as the partition function of the system when $N$ is fixed.

From $  Z(z,y) =\frac1{1-zF(y)}$ one can calculate $\langle N \rangle = z \frac{d}{dz} \ln Z(z,y)$ and  $\langle M \rangle = y \frac{d}{dy} \ln Z(z,y)$ and set  $\langle N \rangle + \langle M \rangle$ to a desired value of $L$  to eliminate $z.$  Particle density $\rho(y)= \frac{ \langle N \rangle}L$ in GCE  is then, 
\be \label{eq:rhoy_app}
\rho(y)=
\frac{1}{1 +y \frac{F'(y)}{F(y)}} =\frac{(1-y)(1- \gamma y) (2-y- \gamma y)}{ (1- \gamma y)^2 + (1-y)^2}. 
\ee

To verify  if MPSS  obtained here is indeed a good  approximation let  us calculate  and compare from Monte Carlo simulations,  the steady state  values  of $\eta_+,$  the average number of beads  per  $+$ urn and $\rho_+,$ the fraction of urns  having internal degree $+,$
\bea \label{eq:eta}
\eta_+ =\frac1N \sum_{k=1}^N \langle m_k \delta_{\sigma_k,+}\rangle;
\rho_+=\frac1N \sum_{k=1}^N  \langle \delta_{\sigma_k,+}\rangle.
\eea
Since simulations are done at some specific $L,N,$  we can use $F(y)^N$ as the partition function of the system; thus
$p_+(m) = y^m/F(y)$  and  $p_-(m) = \gamma^m y^m/F(y)$ and,  
\bea \label{eq:etapm}
\eta_+ &=& \frac1{F(y)}\sum_{m=0}^\infty m Tr[X_+(m)] y^m= \frac{y(1-\gamma y)}{(1-y) (2-y-\gamma y)};\cr
\rho_+ &=& \frac1{F(y)}\sum_{m=0}^\infty Tr[X_+(m)] y^m=\frac{1-\gamma y}{2-y-\gamma y}.
\eea
Using  density-fugacity relation  (\ref{eq:rhoy_app}), both $\eta_+$ and $\rho_+$ can be obtained for different $\rho.$ 

In Fig. \ref{Fig:approx} we plot $\eta_+$ and $\rho_+$ as a function  of $\gamma$ (dashed  lines), for different $\rho$ in the range $(0.1,0.9),$  along with those obtained from the   Monte Carlo simulations of the model (solid lines). They match quite well for all $\gamma <1$, indicating that,  the approximate MPSS  describes the RTP  model very well.
\vspace*{.5 cm}

{\it Acknowledgement:}
PKM  acknowledges stimulating  discussions with Urna Basu.   IM acknowledges the support of Council of Scientific and Industrial Research, India  (Research Fellowship, 
Grant No. 09/921(0335)/2019-EMR-I).

\end{document}